\documentclass[a4paper]{jpconf}
\usepackage{amsmath}
\usepackage{amssymb}
\usepackage{graphicx}

\begin{document}

\newcommand{\HFBAX}{\sc axialhfb}
\newcommand{\HFBTHO}{\sc hfbtho}

\title{Reflection-asymmetric nuclear deformations within the Density Functional Theory}
\author{E Olsen$^1$, J Erler$^{1,2}$, W Nazarewicz$^{1-3}$ and M Stoitsov$^{1,2}$}
\address{$^1$ Department of Physics and Astronomy, University of Tennessee, Knoxville, Tennessee 37996, USA}
\address{$^2$ Physics Division, Oak Ridge National Laboratory, Post Office Box 2008, Oak Ridge, Tennessee 37831, USA}
\address{$^{3}$Institute of Theoretical Physics, Warsaw University, ul. Ho\.{z}a 69, PL-00681, Warsaw, Poland}
\ead{eolsen1@utk.edu}

\begin{abstract}

Within the nuclear density functional theory (DFT)  we study the effect of reflection-asymmetric shapes on ground-state binding energies and binding energy differences.  To this end, we developed the new DFT solver {\HFBAX} that uses an approximate second-order gradient to solve the Hartree-Fock-Bogoliubov equations of superconducting DFT with the quasi-local Skyrme energy density functionals.  Illustrative calculations are carried out for even-even isotopes of radium and thorium.  

\end{abstract}

\section{Introduction}

\begin{figure}[htb]
   \centering
   \includegraphics[width=0.8\textwidth]{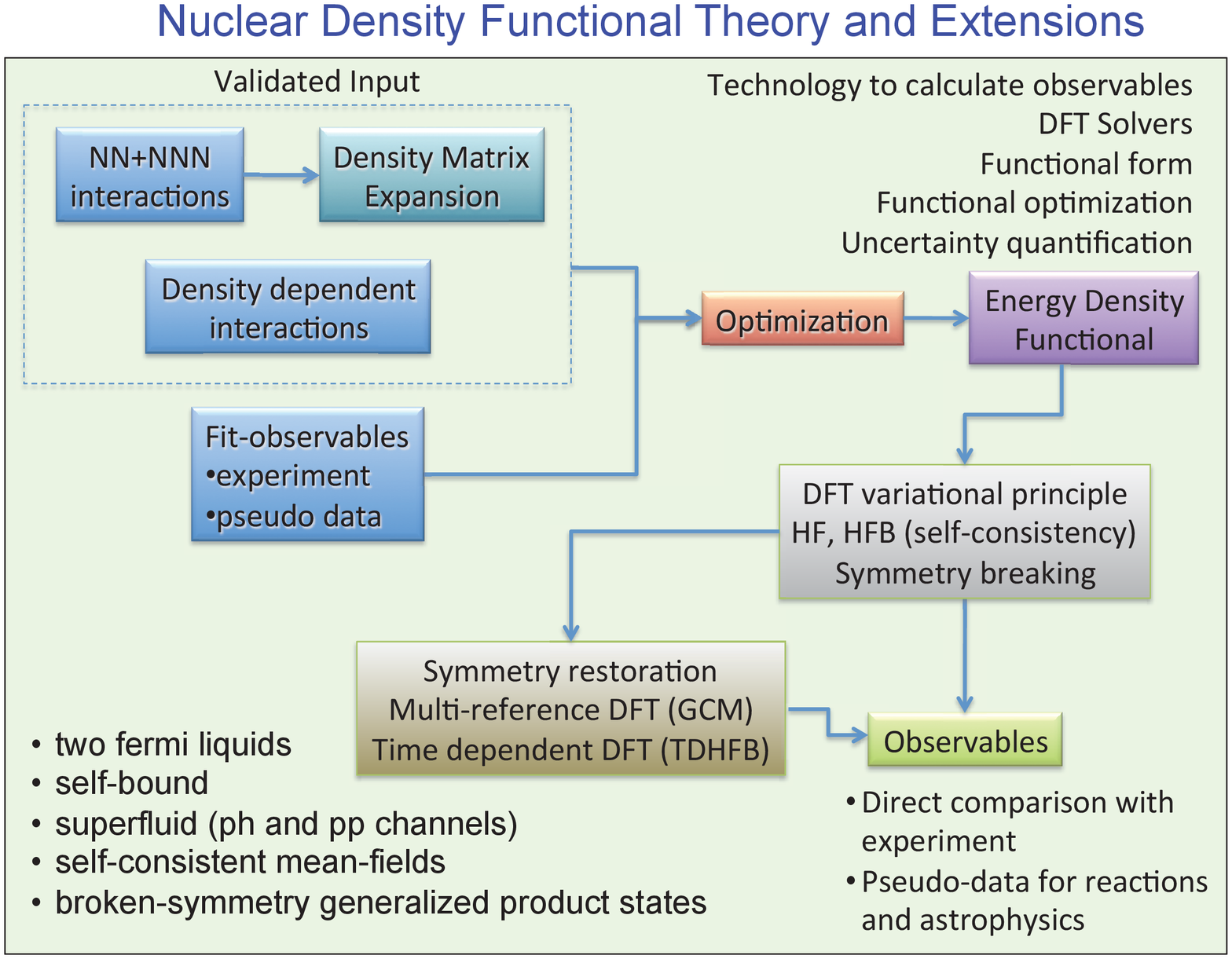}
   \caption{The nuclear DFT strategy diagram.}
   \label{dftstrategy}
\end{figure}

The goal of low-energy nuclear theory  is to provide a comprehensive theoretical framework  to explain properties of atomic  nuclei and their reactions at the nucleonic level. The roadmap towards this goal involves three theoretical strategies: ab initio methods, configuration-interaction techniques, and the nuclear DFT \cite{Ber07,Fur11}.  Due to exploding dimensions of the configuration space, properties of complex heavy nuclei with many valence particles are best described by the nuclear DFT and its various extensions \cite{ref3}. The main idea of DFT is to describe an interacting system of fermions via its densities rather than the many-body wave function.  The energy of the many body system can be written as 
a density functional, and the ground state energy is obtained through the variational procedure.  Figure~\ref{dftstrategy} illustrates the strategy behind the nuclear DFT. A major thrust in this area is to enhance the predictive power of DFT by firmly connecting the nuclear energy functional (EDF) to the nuclear interactions and to optimize the  low-energy coupling constants of the microscopically based EDF by comparing results of theoretical simulations with a selected set of observables \cite{Kor10,Kor11}. 

The 
nuclear DFT is based on the self-consistent mean-field approach rooted
in the self-consistent Hartree-Fock (HF) or Hartree-Fock-Bogoliubov (HFB) problem.  The self-consistent HFB equations allow us to compute
the nuclear ground state and a set of elementary quasi-particle
excitations that can be used to build excitations of the system. The
HFB equations constitute a system of coupled integro-differential
equations that can be written in a matrix form as an eigenvalue
problem, where the dependence of the HFB matrix elements on the
eigenvectors induces nonlinearities. 

Since all the nucleons are
treated on equal footing 
and contribute
to  single-particle and  pairing mean fields, a few hundred nucleonic wave
functions have to be considered which makes systematic calculations
quite demanding.
We have at our disposal a suite
of optimized HFB solvers working in various geometries \cite{Pei08}.
All our DFT solvers adopt an iterative approach to solving the HFB
equations \cite{Ben05,Sto05,Sch11,Fan09,Pei11,Pei12}. Nuclear configurations
(characterized by shape deformations, angular momentum, temperature,
single-particle occupations, etc.) can be selected by means of
external constraints \cite{Sta10a}. By employing modern DFT infrastructure
global nuclear properties 
 can be rapidly computed across the nuclear landscape \cite{Erl12}.

One of the fundamental properties of an atomic nucleus is its shape. The DFT description of nuclei is performed in the frame of
reference of  the nucleus, the intrinsic frame, in which
the nucleus may acquire a deformed shape. The concept of 
shape deformation is ultimately
related to the spontaneous symmetry breaking effect
known in many areas
of physics. 
Symmetry breaking solutions may appear variationally when, in a
mean-field configuration respecting the original symmetries of the nuclear
Hamiltonian, degenerate single-particle orbits are strongly
coupled to collective vibrations \cite{Rei84,Naz92,Naz94,Fra01}. In 
heavy nuclei, the
most important collective modes are the quadrupole and octupole vibrations, leading
to  intrinsic mean fields with non-zero quadrupole and octupole moments.
If a nucleus has nonzero quadrupole moments, its non-spherical shape can be characterized by ellipsoidal deformations \cite{Boh75,ref2}. The presence of octupole moments is indicative of  reflection-asymmetric deformations \cite{ref1}. While quadrupole deformations are common in ground states across the mass table, octupole deformations are more concentrated in particular regions of the chart of the nuclides \cite{ref1,Ahm93}.  

This paper introduces a new DFT solver  {\HFBAX} which solves the HFB problem by means of an approximate second-order gradient method \cite{RobPRC84}.  Using {\HFBAX} with Skyrme EDFs, we  study the effect of reflection-asymmetric moments on nuclear binding energy  in even-even Ra and Th nuclei.  These nuclei were chosen for their well known strong  octupole correlations.  The binding energies obtained in the nuclear DFT are further used to compute mass filters such as two-nucleon separation energies and the double-difference indicator $\delta V_{pn}$ \cite{Zha89,Cak05,ref8}.      
 
This paper is structured as follows.  In Sec.~\ref{secHFB} we give  a brief overview of HFB equations, introduce  the new DFT solver  {\HFBAX},  and provide  necessary details about our  DFT calculations.  The results for the mass filters are presented in Sec.~\ref{secfilters}.  Finally, conclusions and prospects for future work are given in Sec.~\ref{seccon}.

\section{DFT Calculations}\label{secHFB}

The following is a succinct description of the HFB approach.  For a more in-depth discussion, the reader is referred to Refs.~\cite{ref3,ref2}.
The goal is to find the ground state energy of a  nucleus ($N,Z$) subject to constraints.  The energy functional
 is a three-dimensional spatial integral
of local energy density ${\cal H}$ that is a real, scalar,
time-even, and isoscalar function of local densities and their 
derivatives. The energy density  consists of a kinetic term, 
the Skyrme energy functional representing the effective nuclear interaction between nucleons, and
the Coulomb term.  
A Skyrme functional  depends on a number of local densities:
nucleonic  densities, kinetic densities, spin densities,
spin-kinetic densities, current  densities,
tensor-kinetic densities, and spin-current densities \cite{Per04}. Since 
pairing correlations are considered, the EDF also includes pairing terms defined in terms of pairing densities. Variation of the binding energy leads to the HFB equations:
\begin{equation} \label{eq:HFB}
   \left( \begin{array}{cc}
          h - \lambda & \Delta \\
          -\Delta^* & -h^* + \lambda
          \end{array} \right)
   \left( \begin{array}{cc}
           U_k \\
           V_k
           \end{array} \right)
            =  E_k \left( \begin{array}{cc}
                     U_k \\
                     V_k
                  \end{array} \right),
\end{equation}
where  $h$ is the HF Hamiltonian containing the self-consistent field, $\Delta$ is the pairing field, $\lambda$ is the Fermi level, $E_k$ are  one-quasiparticle energy eigenstates, and ($U_k, V_k$) are two-component HFB eigenvectors. To restore the particle symmetry spontaneously broken in the HFB approach  we use the Lipkin-Nogami scheme \cite{Nog64a,Lip60a} implemented as in \cite{Sto07}.

The present calculations were done using a parity-breaking axial DFT solver {\HFBAX} recently developed by Stoitsov using  the gradient method routine for solving the HFB equations (\ref{eq:HFB}) provided by L.M. Robledo \cite{RobPRC84}.  Like in the parity-conserving axial HFB code {\HFBTHO} \cite{Sto05},   {\HFBAX} expresses the HFB equations  in a large configuration space of single-particle eigenstates of the deformed harmonic oscillator basis.    Unlike {\HFBTHO}, however, which employs the  direct diagonalization technique,
{\HFBAX} uses an approximate second-order gradient method \cite{RobPRC84,RobNPA594}.  This method is particularly  well suited to deal with multiple constraints \cite{ref2}; hence,  it does not require special techniques such as the augmented Lagrangian method \cite{Sta10a} implemented in {\HFBTHO}.  

\begin{figure}[htb]
   \centering
   \includegraphics[width=\textwidth]{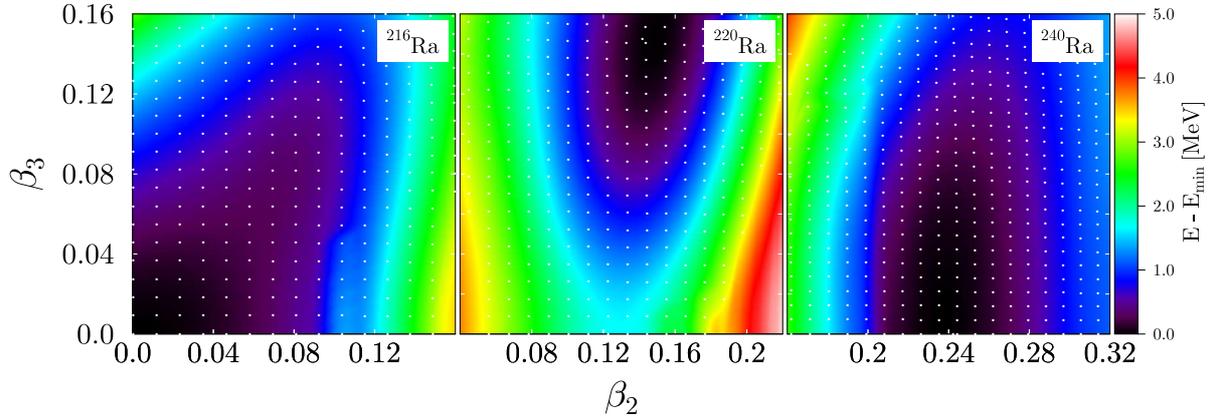}
   \caption{The contour maps of the 2D potential energy surfaces of $^{216}$Ra (left), $^{220}$Ra (middle), and $^{240}$Ra (right) calculated by constraining both the quadrupole and octupole moments characterized by shape deformations $\beta_2$ and $\beta_3$, respectively.  The mesh points at which calculations were performed  are marked by white dots. The energy (in MeV) is shown relative to the ground-state minimum. The Skyrme functional SLy4 \cite{Cha98a} and mixed pairing \cite{DobEPJ1521} were used.
   }
   \label{fig:energysurface}
\end{figure}

As an individual constrained HFB problem  can be solved on a single processor, the computation of a  potential energy surface (PES) in a space of collective coordinates is embarrassingly parallel (see  Fig.~\ref{fig:energysurface}).  This also makes it possible to produce deformed HFB mass tables -- including reflection asymmetric shapes -- on leadership-class supercomputers such as  JAGUAR at Oak Ridge National Laboratory's Leadership Computing Facility.

\begin{figure}[htb]
	\centering
	\includegraphics[width=\textwidth]{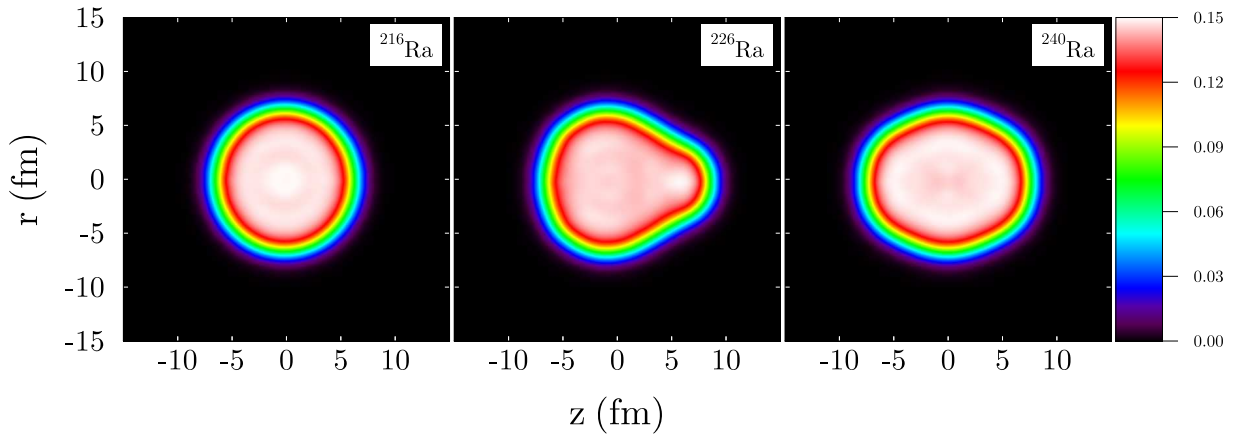}
	\caption{Nucleon density distributions (in cylindrical coordinates $r$ and $z$) calculated for $^{216}$Ra ($\beta_2 = 0.00$, $\beta_3 = 0.00$), $^{226}$Ra ($\beta_2 = 0.20$, $\beta_3 = 0.198$), and $^{240}$Ra ($\beta_2 = 0.24$, $\beta_3 = 0.0$)   
	showing spherical, quadrupole and octupole deformed shapes. 
	}
	\label{fig:densityplots}
\end{figure}
The goal was to find the ground state (g.s.) energy, i.e. the global minimum on a 2D potential energy surface.  To guarantee that a global, rather than local,   minimum is found, one has to compute and scan  the PES  in two coordinates:  the axial mass quadrupole moment $Q_{20}$ (characterizing the elongation of the nucleus expressed in terms of deformation $\beta_2$) and the mass octupole moment $Q_{30}$ (characterizing the mirror asymmetry, or pear-like deformation $\beta_3$).  Figure \ref{fig:energysurface} shows three energy surfaces for different isotopes of Radium  calculated using the Skyrme EDF SLy4 \cite{Cha98a} together with a density dependent mixed pairing functional \cite{DobEPJ1521}.  All of the calculations here were done using a large oscillator basis of 21 oscillator shells (2024 basis states).  
We performed 525 constrained calculations for different combinations of $Q_{20}$ and $Q_{30}$ (indicated by white dots on the plots in Fig.~\ref{fig:energysurface}). The range of $Q_{30}$-values was identical for each $Q_{20}$; they went from $0$ to 6000\,fm$^3$ in steps of 250\,fm$^3$. 
Following the minimization on the resulting grid,  unconstrained calculations have been carried out starting from the grid minimum: in this way the precise total g.s. binding energy was obtained and used to compute mass filters.

Figures~\ref{fig:energysurface} and  \ref{fig:densityplots} illustrate three  situations typical to all nuclei considered \cite{Naz84,Rob87}. In general,  for a given isotopic chain, a transition is expected with increasing neutron number from spherical shape to quadrupole well-deformed shapes, via reflection-asymmetric minima.
The PES for $^{216}$Ra is characteristic
of a spherical system, which is fairly soft in the quadrupole-octupole direction. For  $^{220}$Ra, our calculations yield a well-developed minimum with stable quadrupole and octupole deformations. As seen in Fig.~\ref{fig:densityplots}, in  its ground state  $^{226}$Ra is predicted to have a pear-like shape. The nucleus
$^{240}$Ra is strongly deformed but reflection-symmetric. Its shape is indicative of appreciable quadrupole and hexadecapole  g.s. moments.

\section{Results}\label{secfilters}

The objective of this work was to determine whether reflection-asymmetric g.s. moments   may have an affect on binding energy differences (mass filters) of even-even Ra and Th isotopes.
\begin{figure}[htb]
	\centering
\includegraphics[width=1.0\textwidth]{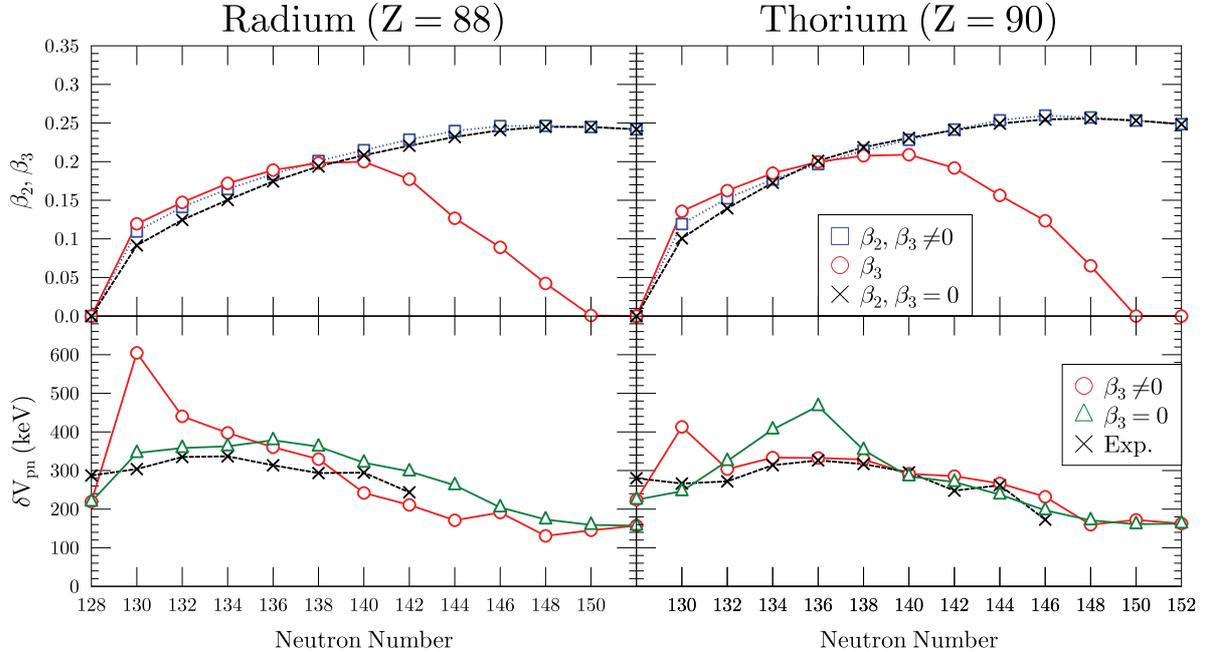}
	\caption{Results of HFB calculations using the EDF SLy4 \cite{Cha98a} and mixed pairing  \cite{DobEPJ1521} for the even-even isotopes of Ra and Th.  Top: g.s. deformation parameters $\beta_2$ and $\beta_3$ obtained in the full minimization allowing reflection-asymmetric shapes ($\beta_3 \ne 0$) compared with those obtained by assuming mirror symmetry ($\beta_3=0$).   Bottom:  $\delta V_{pn}$  for reflection-symmetric (triangles) and asymmetric (circles) shapes compared with   experimental values  ($\times$) of Ref.~\cite{Audi2003337}.}
	\label{fig:alldata1}
\end{figure}

Figure~\ref{fig:alldata1} (top) shows the equilibrium deformations $\beta_2$ and $\beta_3$ predicted with the Skyrme functional SLy4. It is seen that the octupole deformation maximizes around $N$=138 and then gradually decreases with the neutron number.  The effect of the octupole deformation on the quadrupole moment appears to be very small. This is evident by comparing the g.s. $\beta_2$-values obtained in the full minimization ($\beta_3 \ne 0$) with those obtained by assuming reflection symmetric shapes ($\beta_3=0$).
The bottom part of Fig.~\ref{fig:alldata1} shows the filter
$\delta V_{pn}$:
	\begin{equation}
  	\delta V_{pn} = \frac{1}{4} \left[B(Z,N) - B(Z,N-2) - B(Z-2,N) + B(Z-2,N-2) \right],
  \end{equation} 
which can be viewed as an approximation to the  mixed partial derivative \cite{ref8}:
\begin{equation}
	\delta V_{pn}(Z,N) \approx \frac{\partial^2 B}{\partial Z \partial N}.
\end{equation}
In general, the agreement with experiment for $\delta V_{pn}$ is improved   when octupole correlations are considered.   The  spike around $N=130$ in the $\beta_3 \ne 0$ results is  due to the  rapid transition between spherical and deformed shapes; this effect -- typical to mean-field calculations -- is supposed to be washed out if beyond-mean-field effects are taken into account.

\begin{figure}[htb]
	\centering
\includegraphics[width=1.0\textwidth]{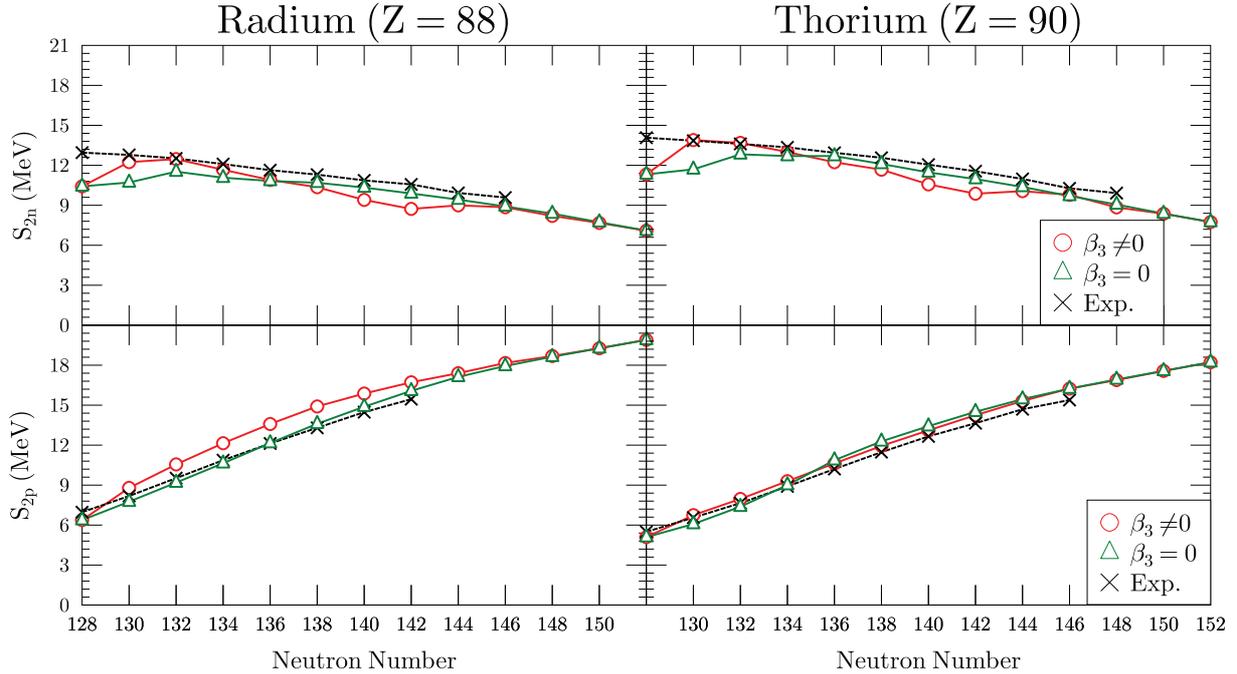}
	\caption{Similar to Fig.~\ref{fig:alldata1} except for two-neutron (top) and two-proton (bottom) separation energies.
	}
	\label{fig:alldata2}
\end{figure}

Finally, Fig.~\ref{fig:alldata2} shows predicted and experimental  two-neutron and two-proton separation energies:
  \begin{equation}
     S_{2n} = B(Z,N) - B(Z,N-2),~~~S_{2p} = B(Z,N) - B(Z-2,N).
  \end{equation}
For these mass filters, the  effect of octupole correlations is very small.

\section{Conclusions}\label{seccon}

The large-scale simulations utilizing  leadership-class supercomputers have transformed the DFT-based description  of nuclei. A close coupling
of math and computer scientists with physicists, focused on a close
coupling with present and future experiments and observations, has
dramatically advanced the field \cite{Nam12}.

This work presents selected results of large-scale DFT calculations including reflection-asymmetric shape degrees of  freedom. We focus on the influence of parity-breaking intrinsic moments on the binding energy differences.
Our calculations suggest that
 $\delta V_{pn}$ is  affected by octupole effects, bringing it closer to experiment, whereas $S_{2n}$ and $S_{2p}$ are not. 
 Future work will involve performing similar calculations across the nuclear chart, also for odd-$A$ nuclei.  This will require extending {\HFBAX} to the odd-particle case \cite{Sch10}. We also intend to study the model dependence of our predictions by considering different energy density functionals.

\section*{Acknowledgments}

Discussions with Rick Casten and Burcu Cakirli are gratefully acknowledged.  In particular we want to thank Luis Robledo for providing the  conjugate gradient routine.  This work was supported by the U.S. Department of Energy under Contract Nos.\  DE-FG02-96ER40963 (University of Tennessee) and DE-FC02-09ER41583 (UNEDF SciDAC Collaboration).  Computational resources were provided through an INCITE award ``Computational Nuclear Structure'' by the National Center for Computational Sciences (NCCS) and National Institute for Computational Sciences (NICS) at Oak Ridge National Laboratory.

\section*{References}

\bibliography{references2}

\providecommand{\newblock}{}
\begin{thebibliography}{10}
\expandafter\ifx\csname url\endcsname\relax
  \def\url#1{{\tt #1}}\fi
\expandafter\ifx\csname urlprefix\endcsname\relax\def\urlprefix{URL }\fi
\providecommand{\eprint}[2][]{\url{#2}}

\bibitem{Ber07}
Bertsch G, Dean D and Nazarewicz W 2007 {\em SciDAC Review\/} {\bf 102} 42

\bibitem{Fur11}
Furnstahl R 2011 {\em Nucl. Phys. News\/} {\bf 21}(2) 18

\bibitem{ref3}
Bender M, Heenen P~H and Reinhard P~G 2003 {\em Rev. Mod. Phys.\/} {\bf 75} 123

\bibitem{Kor10}
Kortelainen M, Lesinski T, Mor\'e J, Nazarewicz W, Sarich J, Schunck N,
  Stoitsov M~V and Wild S 2010 {\em Phys. Rev. C\/} {\bf 82} 024313

\bibitem{Kor11}
Kortelainen M, McDonnell J, Nazarewicz W, Reinhard P~G, Sarich J, Schunck N,
  Stoitsov M~V and Wild S 2011 {\em Phys. Rev. C\/} {s}ubmitted;
  arXiv:1111.4344

\bibitem{Pei08}
Pei J~C, Stoitsov M~V, Fann G~I, Nazarewicz W, Schunck N and Xu F~R 2008 {\em
  Phys. Rev. C\/} {\bf 78} 064306

\bibitem{Ben05}
Bennaceur K and Dobaczewski J 2005 {\em Comput. Phys. Commun.\/} {\bf 168} 96

\bibitem{Sto05}
Stoitsov M, Dobaczewski J, Nazarewicz W and Ring P 2005 {\em Comput. Phys.
  Commun.\/} {\bf 167} 43

\bibitem{Sch11}
Schunck N, Dobaczewski J, McDonnell J, Satu{\l}a W, Sheikh J, Staszczak A,
  Stoitsov M and Toivanen P 2012 {\em Comput. Phys. Commun.\/} {\bf 183} 166

\bibitem{Fan09}
Fann G, Pei J, Harrison R, Jia J, Hill J, Ou M, Nazarewicz W, Shelton W and
  Schunck N 2009 {\em J. Phys. Conf. Ser.\/} {\bf 180} 012080

\bibitem{Pei11}
Pei J~C, Kruppa A~T and Nazarewicz W 2011 {\em Phys. Rev. C\/} {\bf 84} 024311

\bibitem{Pei12}
Pei J, Fann G, Harrison R, Nazarewicz W, Hill J, Galindo D and Jia J 2012 {\em
  J. Phys. Conf. Ser.,\/} {this volume}

\bibitem{Sta10a}
Staszczak A, Stoitsov M, Baran A and Nazarewicz W 2010 {\em Eur. Phys. J. A\/}
  {\bf 46} 85

\bibitem{Erl12}
Erler J, Birge N, Kortelainen M, Nazarewicz W, Olsen E, Perhac A and Stoitsov M
  2012 {\em J. Phys. Conf. Ser.,\/} {this volume}

\bibitem{Rei84}
Reinhard P~G and Otten E 1984 {\em Nucl. Phys. A\/} {\bf 420}

\bibitem{Naz92}
Nazarewicz W 1992 {\em Prog. Part. Nucl. Phys.\/} {\bf 28} 307

\bibitem{Naz94}
Nazarewicz W 1994 {\em Nucl. Phys. A\/} {\bf 574} 27c

\bibitem{Fra01}
Frauendorf S 2001 {\em Rev. Mod. Phys.\/} {\bf 73} 463

\bibitem{Boh75}
Bohr A and Mottelson B 1975 {\em Nuclear Structure, vol.\ II\/} (W\ A.\
  Benjamin, Reading)

\bibitem{ref2}
Ring P and Schuck P 1980 {\em The Nuclear Many-Body Problem\/} (New York:
  Springer)

\bibitem{ref1}
Butler P~A and Nazarewicz W 1996 {\em Rev. Mod. Phys.\/} {\bf 68} 351

\bibitem{Ahm93}
Ahmad I and Butler P 1993 {\em Ann. Rev. Nucl. Part. Sci.\/} {\bf 43} 71

\bibitem{RobPRC84}
Robledo L and Bertsch G~F 2011 {\em Phys. Rev. C\/} {\bf 84} 014312

\bibitem{Zha89}
Zhang J~Y, Casten R and Brenner D 1989 {\em Phys. Lett. B\/} {\bf 227} 1

\bibitem{Cak05}
Cakirli R~B, Brenner D~S, Casten R~F and Millman E~A 2005 {\em Phys. Rev.
  Lett.\/} {\bf 94} 092501

\bibitem{ref8}
Stoitsov M, Cakirli R~B, Casten R~F, Nazarewicz W and Satula W 2007 {\em Phys.
  Rev. Lett.\/} {\bf 98} 132502

\bibitem{Per04}
Perli\'nska E, Rohozi\'nski S~G, Dobaczewski J and Nazarewicz W 2004 {\em Phys.
  Rev. C\/} {\bf 69} 014316

\bibitem{Nog64a}
Nogami Y 1964 {\em Phys. Rev. B\/} {\bf 134} 313

\bibitem{Lip60a}
Lipkin H~J 1960 {\em Ann. Phys., NY\/} {\bf 9} 272

\bibitem{Sto07}
Stoitsov M~V, Dobaczewski J, Kirchner R, Nazarewicz W and Terasaki J 2007 {\em
  Phys. Rev. C\/} {\bf 76} 014308

\bibitem{RobNPA594}
Egido J~L, Lessing J, Martin V and Robledo L~M 1995 {\em Nucl. Phys. A\/} {\bf
  594} 70

\bibitem{Cha98a}
Chabanat E, Bonche P, Haensel P, Meyer J and Schaeffer R 1998 {\em Nucl. Phys.
  A\/} {\bf 635} 231, {\textit{{N}ucl. {P}hys. A}} \textbf{643}, 441(E)

\bibitem{DobEPJ1521}
Dobaczewski J, Nazarewicz W and Stoitsov M~V 2002 {\em Eur. Phys. J. A\/} {\bf
  15} 21

\bibitem{Naz84}
Nazarewicz W, Olanders P, Ragnarsson I, Dudek J, Leander G, M{\"o}ller P and
  Ruchowska E 1984 {\em Nucl. Phys. A\/} {\bf 429} 269

\bibitem{Rob87}
Robledo L, Egido J, Berger J and Girod M 1987 {\em Phys. Lett. B\/} {\bf 187}
  223

\bibitem{Audi2003337}
Audi G, Wapstra A and Thibault C 2003 {\em Nucl. Phys. A\/}  337

\bibitem{Nam12}
Nam H, Stoitsov M, Nazarewicz W, Bulgac A, Hagen G, Kortelainen M, Maris P, Pei
  J, Roche K, Schunck N, Thompson I, Vary J and Wild S 2012 {\em J. Phys. Conf.
  Ser.,\/} {this volume}

\bibitem{Sch10}
Schunck N, Dobaczewski J, McDonnell J, Mor\'e J, Nazarewicz W, Sarich J and
  Stoitsov M~V 2010 {\em Phys. Rev. C\/} {\bf 81} 024316

\end{thebibliography}

\bibliographystyle{iopart-num}

\end{document}